\documentclass[final,1p,times]{elsarticle}
\usepackage{graphicx}         
\usepackage[english]{babel}
\usepackage{geometry}
\usepackage{amsmath}
\usepackage{array}
\usepackage{helvet}
\usepackage{multicol}
\usepackage{setspace}
\usepackage{color}
\usepackage{eurosym}
\usepackage{natbib}
\usepackage[T1]{fontenc}
\usepackage[utf8]{inputenc}
\usepackage{epstopdf}
\DeclareGraphicsExtensions{.jpg,.eps,.png,.pdf}
\usepackage{lineno}

\begin{document}
\title{\large{Inclusive J/$\psi$ and $\psi$(2S) production in p-Pb collisions\\
 at $\sqrt{s_{\mathrm{NN}}}$ = 5.02~TeV with ALICE at the LHC}} 

\author[pi]{M.~Winn}
\author{on behalf of the ALICE Collaboration}

\address[pi]{Physikalisches Institut, Universität Heidelberg}

\selectlanguage{english}

\begin{abstract}
We report on the inclusive J/$\psi$ nuclear modification factor in p--Pb collisions at $\sqrt{s_{\mathrm{NN}}}~=~5.02$~TeV as a function of rapidity $y$ 
 and transverse momentum $p_{\mathrm{T}}$. The experimental coverage extends down to $p_{\mathrm{T}}=0$~GeV/$c$ in the three rapidity ranges accessible by ALICE ($-4.46~<~y_{\mathrm{cms}}~< -2.96$,  $-1.37~<~y_{\mathrm{cms}}~<~0.46$,  $2.03~<~y_{\mathrm{cms}}~<~3.53$). The obtained results as a function of rapidity  are in agreement with theory predictions  based only on shadowing or on coherent energy loss. 
At forward and backward rapidity, the $\psi$(2S) measurement complements the J/$\psi$ results. The  ratio between the $\psi$(2S) and J/$\psi$ cross section  is significantly smaller in p--Pb than in pp collisions in both rapidity regions.
\end{abstract}
\maketitle

The seminal paper from T.~Matsui and H.~Satz~\cite{jpsiqgpmatsuisatz} published in 1986 initiated intense theoretical and experimental studies of J/$\psi$ production as a probe of deconfinement in nucleus-nucleus (AA) collisions. 
At LHC energies, in addition to the originally proposed J/$\psi$ suppression due to colour screening,  the production of J/$\psi$ due to (re-)combination of charm quarks is suggested as an additional mechanism in two theoretical approaches: the statistical hadronisation at the confinement phase boundary~\cite{PBMJS_nonthermalaspectsjpsi_2000,Andronic:2007bi} and transport models~\cite{thewsetal_enhancedJpsiProdInDeconfMatter_2001, ZhaoRapp_jpsiregen_2011,YanLiZhuangXuetal_jpsicompregenandsuppr_2006}. The latter assume destruction and production of charmonium states predominantly during the deconfined stage of the system evolution in addition to the primordial production.
 In this context, the investigation of charmonia in p--Pb collisions at the LHC represents a crucial ingredient for the interpretation of J/$\psi$ AA-measurements. 
It  gives access to nuclear effects, which are not attributable to deconfinement, but can alter the nuclear modification factor $R_{\mathrm{AA}} = N_{\mathrm{J/\psi, AA }}/( <T_{\mathrm{AA}}> \cdot \sigma_{ \mathrm{J/\psi, pp}})$.
 $N_{\mathrm{J/\psi, AA}}$ represents in this formula the efficiency corrrected J$/ \psi$-yield per event in AA collisions,  $T_{\mathrm{AA}}$ denotes the nuclear overlap function and $\sigma_{\mathrm{J/\psi, pp}} $ the J/$\psi$ cross section in proton-proton (pp) collisions at the same centre of mass energy per nucleon-nucleon collision.

ALICE has measured J/$\psi$ in p--Pb collisions at a centre of mass energy of $5.02$ TeV in the central barrel detectors  with a pseudorapidity acceptance $|\eta_{\mathrm{lab, track}}|<0.9$  via the dielectron decay channel  and in the forward spectrometer with an acceptance of $2.5<\eta_{\mathrm{lab, muon}}<4.0$   via the dimuon decay channel. A description of the experiment can be found in~\cite{Aamodt:2008zz}. 
 An inversion of the beam direction 
has enabled the muon spectrometer to measure the J/$\psi$ on the proton and on the lead fragmentation side in the asymmetric p--Pb collision system. In the following, the rapidity of the incoming proton is chosen to  be positive, 
which fixes the rapidity sign convention of the dimuon pair. 
The centre of mass system of the two colliding projectiles is boosted with respect to the laboratory frame system by $\Delta y = 0.465$ resulting in slightly asymmetric rapidity coverages with respect to $y = 0$: $-4.46~<~y_{\mathrm{cms}}~<~-2.96$ at backward, $-1.37<y_{\mathrm{cms}}<0.43$ at central rapidity and $2.03<y_{\mathrm{cms}}<3.53$ at forward rapidity. In the dimuon acceptance,  the J/$\psi$ production was measured in six equally large ranges of $y_{\mathrm{cms}}$ in both accessible regions integrated over transverse momentum. The transverse momentum dependence at forward and backward rapidity was investigated  in seven intervals ($0$-$1$, $1$-$2$, $2$-$3$, $3$-$4$, $4$-$5$,$5$-$6$, $6$-$8$~GeV/$c$). At central rapidity, the signal yield was extracted integrated over $p_{\mathrm{T}}$ and $y_{\mathrm{cms}}$  and in five intervals of transverse momentum ($0$-$1.3$, $1.3$-$3.0$, $3.0$-$5.0$, $5.0$-$7.0$ and $7.0$-$10.0$ GeV/$c$).
The presented results rely 
 in case of the dielectrons ($L_{\mathrm{int}}= 52~ \mu$b$^{-1}$) on a minimum bias trigger\footnote{Requirement of a coincident signal from two scintillator arrays covering $-5.1<\eta_{\mathrm{lab}}<-2.8$ (lead fragmentation side) and $1.7<\eta_{\mathrm{lab}}<3.7$ (the proton fragmentation side).}  and on a dimuon trigger ($L_{\mathrm{int}}= 5.0 (5.8) $~nb$^{-1}$) at forward (backward) rapidity in case of the forward spectrometer.

The signal extraction in the dimuon analysis is based on a Crystal Ball fit with asymmetric tails or pseudo-Gaussian phenomenological shapes for the signal and a variable width Gaussian or the product of a $4^{\mathrm{th}}$-order polynomial and an exponential for the background.  The systematic uncertainty on the signal extraction is estimated from the Root Mean Square of the extracted J/$\psi$ yields obtained combining different signal and background shape assumptions and varies between $1$-$4\ \%$ depending on $p_{\mathrm{T}}$ and rapidity.

The analysis at central rapidity uses electron candidate tracks within $|\eta_{\mathrm{lab}}|<~$0.9. The measurement relies on the particle identification provided by the ALICE Time Projection Chamber via the specific energy loss in the counting gas. The signal extraction uses a background constructed from the like-sign pairs, scaled to match the integral of the opposite-sign dielectron invariant mass distribution in the range 3.2-4.9 GeV/$c^2$. As a  second approach to estimate the systematic uncertainty, a fit to the invariant mass distribution was employed, where the background shape is described by a ratio of polynomials and the signal shape is extracted from simulations with full transport and detector response. In both approaches, the signal yields are obtained by bin counting after subtraction of the background component in the invariant mass range $2.92-3.16$ GeV/$c^2$. The correction for the non-negligible fraction of signal counts, which are reconstructed with lower invariant masses than 2.92 GeV/$c^2$ due to the radiative decay contribution by J$/\psi \to e^+ e^- \gamma$ and bremsstrahlung of the decay particles, is computed using simulations and amounts to $ 31 \ \%$.
 
\begin{figure}
\centering
\begin{minipage}{0.5 \textwidth}
\includegraphics[ width=1.0 \textwidth]{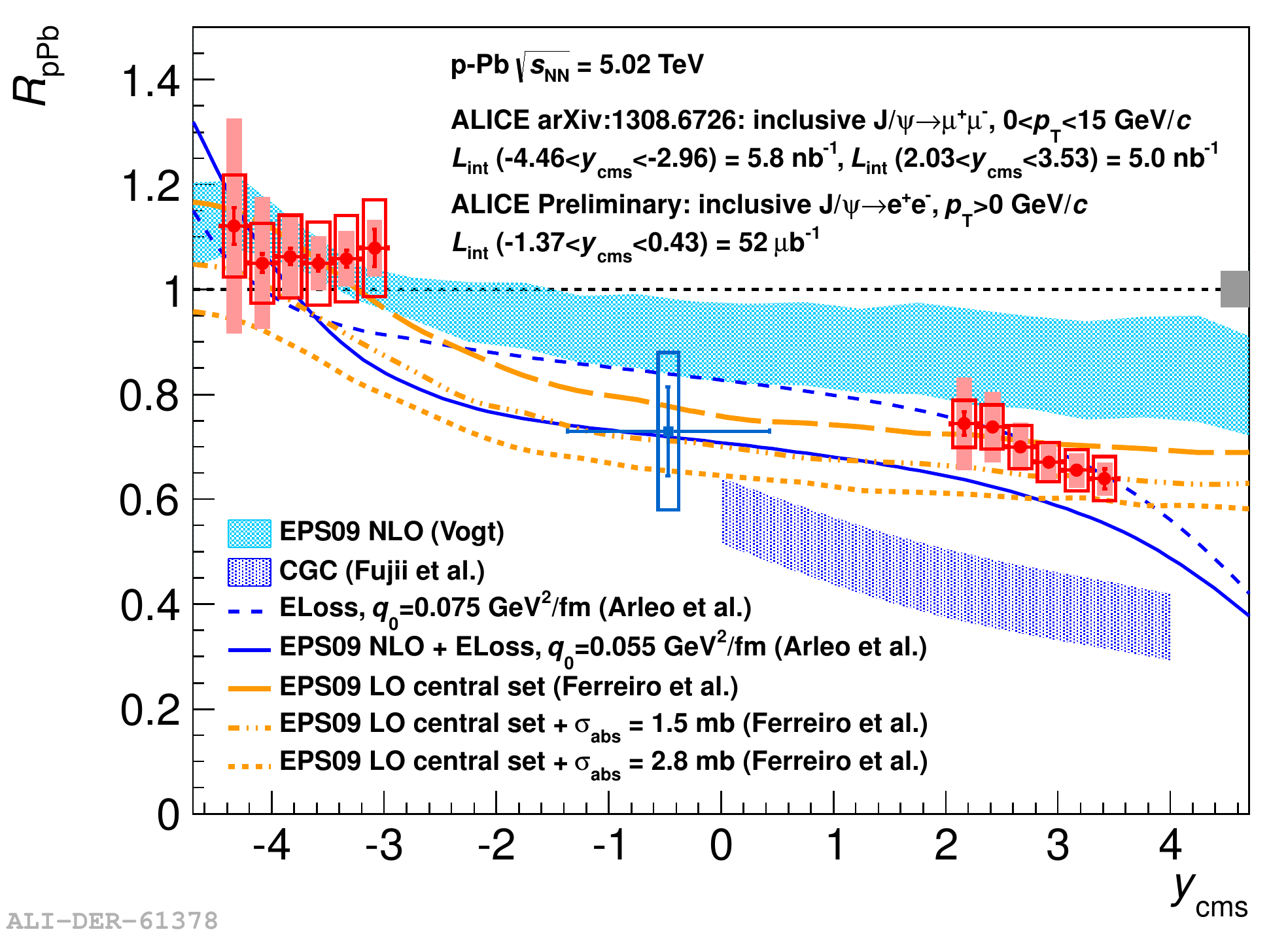}
\end{minipage}
\begin{minipage}{0.49 \textwidth}
\caption{The nuclear modification of inclusive J/$\psi$ as a function of rapidity compared with theory predictions. The open boxes indicate the uncorrelated systematical uncertainties, the filled areas the partially correlated systematical uncertainties and the grey box at unity the common normalization uncertainty due to the $T_{\mathrm{pA}}$-uncertainty. The theory predictions can be found in~\cite{Albaceteetal_predictionspPbatLHC_2013_RVogtversion,FerreiroFleuretLansberg_Rakotozafindrabe_jpsi_pPb_2013, arleo_heavyQuarkoniumSupprInpAFromPartonEnergyLoss_2013, Fujii_Watanabe_pAquarkoniaCGC_2013}.
} 
\label{fig:rpAasafcty}
\end{minipage}

\end{figure}

In case of the results in the dimuon decay channel, the pp-reference for the nuclear modification factor relies on an interpolation of the ALICE measurements at $\sqrt{s} = $ 2.76 TeV and 7 TeV in the muon spectrometer and is performed bin-by-bin in $y$ or $p_{\mathrm{T}}$. For the rapidity bins not covered due to the rapidity shift, an extrapolation is used. More details about the explored method as a function of rapidity can be found in~\cite{LHCb_ALICE_pubnote_pprefforpPb_2013}.

For the dielectron decay channel, the reference cross section was evaluated based on the interpolation of BR~$ \times \textrm{d}\sigma/\textrm{d} y$ measurements in pp(p\={p}) collisions at central rapidity published by the PHENIX Collaboration at $\sqrt{s} = 200$~GeV~\cite{Adare_phenix_jpsi_pp_ref_200GeV_2006},  by the CDF Collaboration at $\sqrt{s} = 1960 $ GeV~\cite{Acosta_CDF_pprefjpsi_1960_2004} and the ALICE Collaboration at $\sqrt{s} = 2.76 $ TeV~\cite{ALICE_jpsi2.76pp_2012} and $\sqrt{s} = 7$~TeV~\cite{Aamodt2011}. 
 At $\sqrt{s}~=~5.02$ TeV, the procedure provides BR $ \times \textrm{d}\sigma/\textrm{d} y~=~367.8 \pm 61 $ nb. The effect of  the slightly shifted rapidity window is negligible compared to the size of the uncertainties.  The pp-reference uncertainty is the largest contribution to the systematic uncertainty on the central rapidity $R_{\mathrm{pA}}$. For the $p_{\mathrm{T}}$-dependent reference, a phenomenological scaling inspired by~\cite{Bossuetal_jpsicrosssectioninterpolationinppforref_2011} was used.

Figure \ref{fig:rpAasafcty} shows the nuclear modification factor $R_{\mathrm{pA}}$ as a function of rapidity integrated over $p_{\mathrm{T}}$ compared to models. The forward and backward results~\cite{ALICE_pPb_firstjpsi_2013} are consistent with those obtained by the LHCb Collaboration~\cite{LHCb_jpsi_pPb_2013}. The result at forward rapidity  indicates a suppression of J/$\psi$, while the backward rapidity data shows no significant suppression within the uncertainties. The central rapidity measurement is consistent with the forward rapidity result albeit exhibiting larger uncertainties.  
  The predictions by R. Vogt employing NLO nPDF EPS09\cite{Eskolaetal_nPDF_EPS09} and the Colour Evaporation Model (CEM)~\cite{Albaceteetal_predictionspPbatLHC_2013_RVogtversion} as well as the calculations by E. Ferreiro et al.~\cite{FerreiroFleuretLansberg_Rakotozafindrabe_jpsi_pPb_2013} with EPS09\cite{Eskolaetal_nPDF_EPS09} at LO assuming $2 \to 2$ production mechanism $(gg \to J/\psi g)$ are consistent with the experimental results. In case of the model by E. Ferreiro, the possible impact of a finite effective nuclear absorption of J/$\psi$ resonance or the precursor state is also calculated and shown. The data can be described within this model without this absorption mechanism.
 The coherent energy loss model~\cite{arleo_heavyQuarkoniumSupprInpAFromPartonEnergyLoss_2013} is also consistent with data, with or without inclusion of shadowing as an additional effect.  The prediction by H.~Fujii et al.~\cite{Fujii_Watanabe_pAquarkoniaCGC_2013} employing the CEM in a  Color Glass Condensate (CGC) framework is instead disfavoured by the forward rapidity data.

\begin{figure}[t]
\centering
\begin{minipage}{.30 \textwidth}
\includegraphics[ width=1.035 \textwidth]{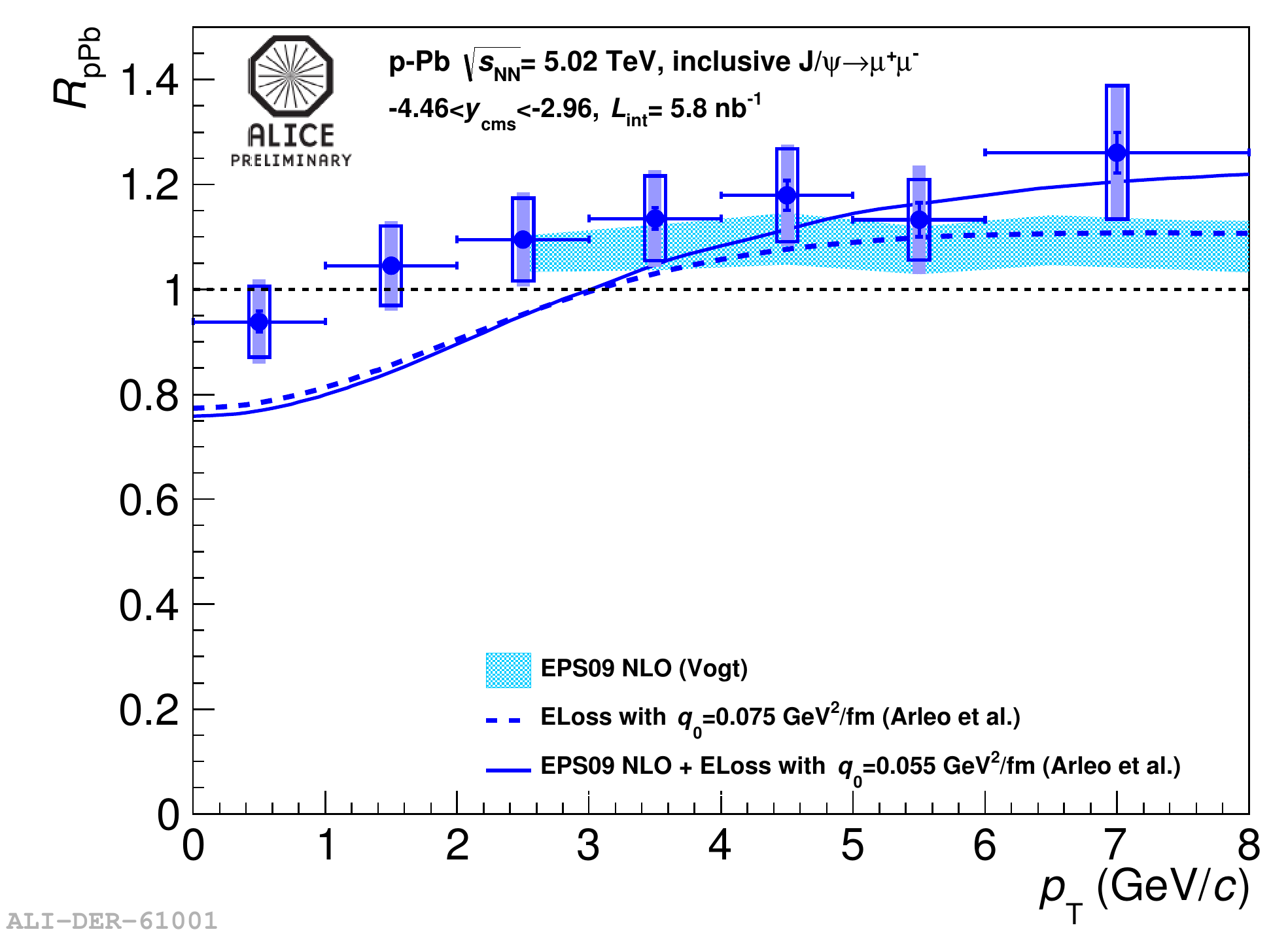}
\end{minipage}
\begin{minipage}{.33 \textwidth}
\includegraphics[ width=1.045 \textwidth]{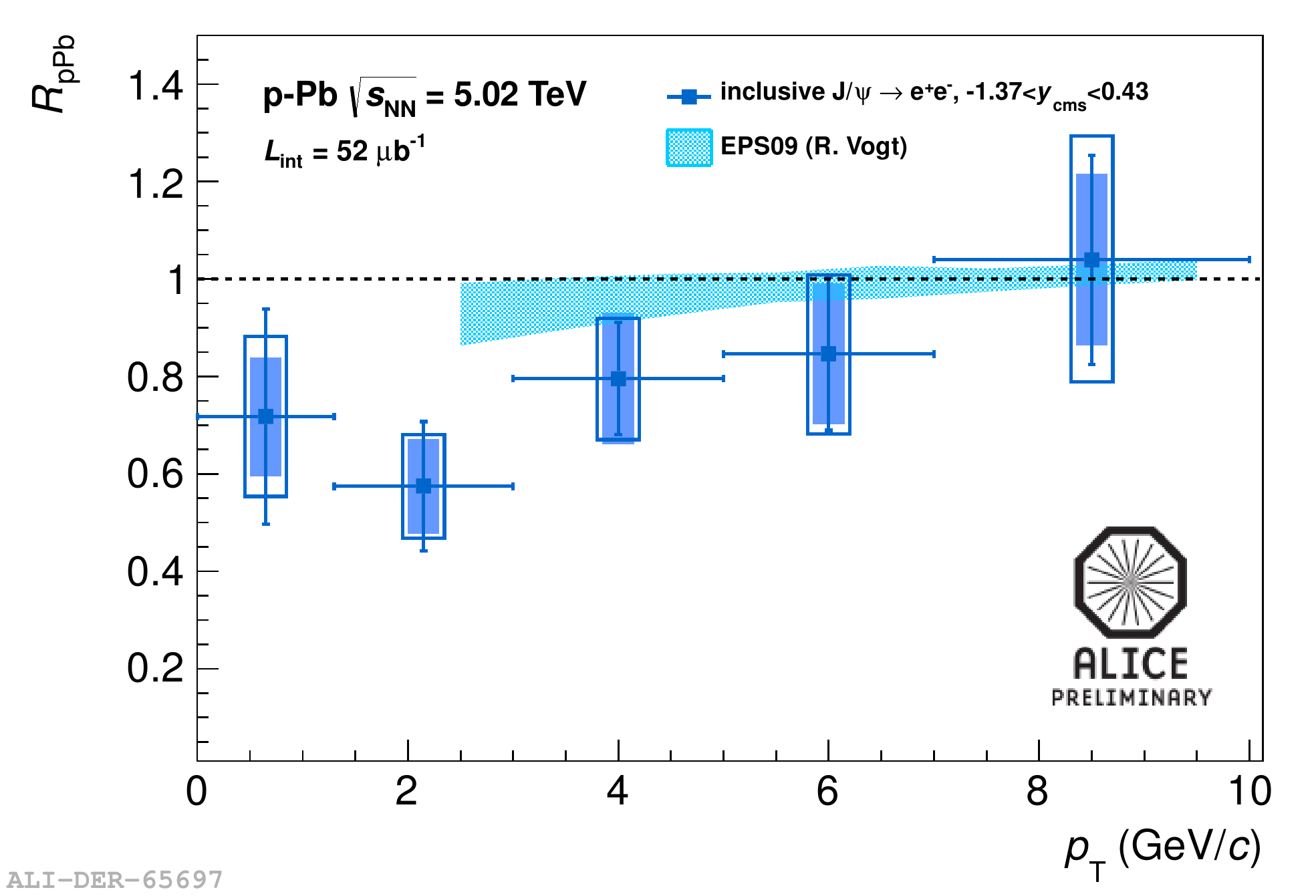}
\end{minipage}
\begin{minipage}{.32 \textwidth}
\includegraphics[ width=.95 \textwidth]{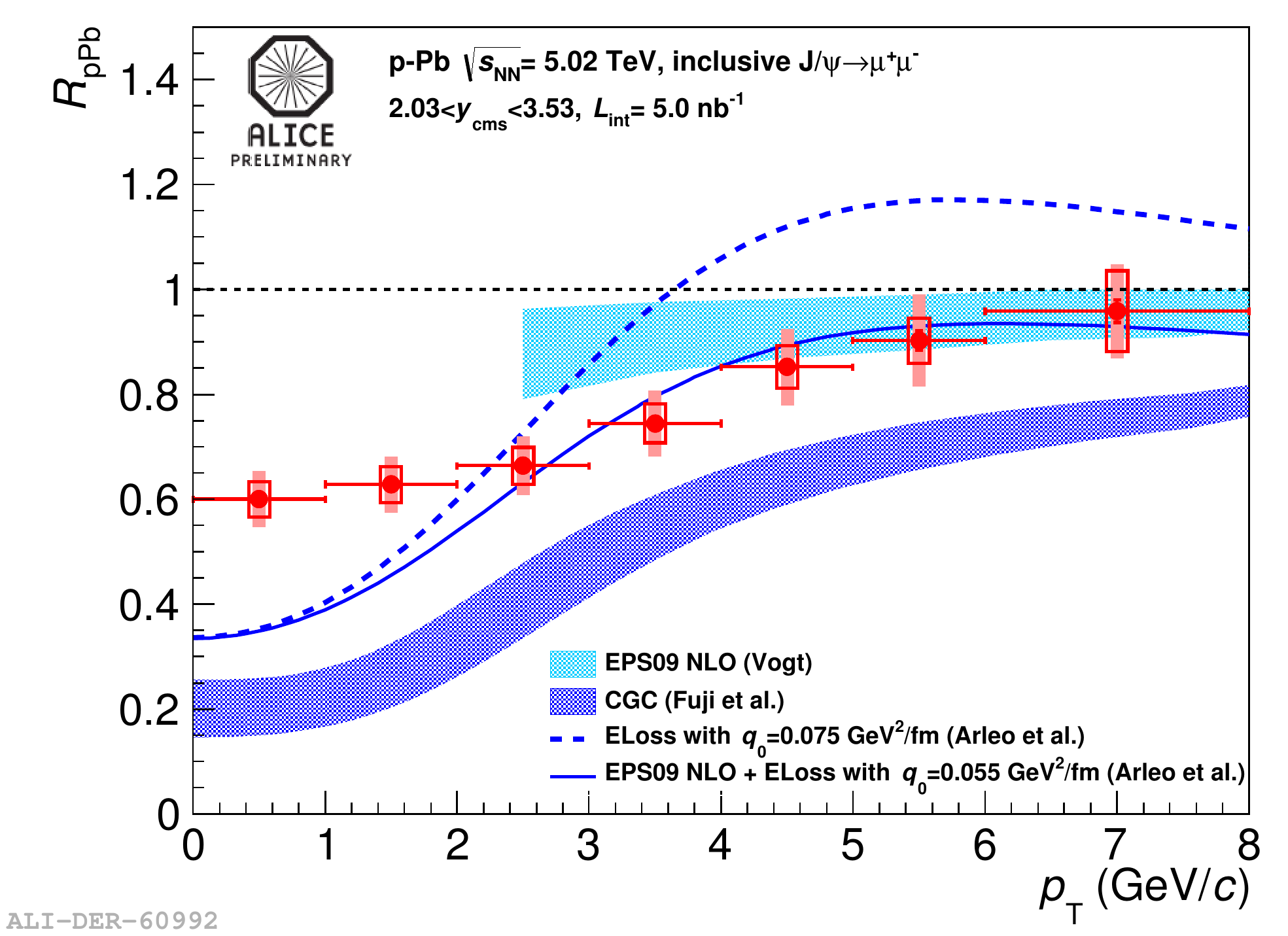}
\end{minipage}

\caption{The J/$\psi$ nuclear modification factor in p--Pb as a function of  $p_{\mathrm{T}}$ at backward rapidity, central rapidity and forward rapidity compared with theory predictions.
The open boxes represent the uncorrelated systematic uncertainties, the filled areas the partially correlated systematical uncertainties. The theory predictions are taken from~\cite{Albaceteetal_predictionspPbatLHC_2013_RVogtversion,Arleo_Peigne_cent_pT_dep_jpsi_suppr_pA_2013,Fujii_Watanabe_pAquarkoniaCGC_2013}.
}
\label{fig:rpAasafctofpt}

\end{figure}

Figure \ref{fig:rpAasafctofpt} shows the $p_{\mathrm{T}}$-dependence of $R_{\mathrm{pA}}$ at backward, mid- and forward rapidity.  At forward rapidity, the suppression present at low $p_{\mathrm{T}}$ is gradually weakening with increasing $p_{\mathrm{T}}$. At high $p_{\mathrm{T}}$, the data is consistent with no suppression. A similar pattern is present at central rapidity, although uncertainties prevent a firm conclusion. At backward rapidity, a feeble $p_{\mathrm{T}}$ dependence seems to be present considering the partially correlated uncertainties with $R_{\mathrm{pPb}}$-values close to unity. 
 The CGC-model for forward rapidity  by H.~Fujii et al.~\cite{Fujii_Watanabe_pAquarkoniaCGC_2013} is disfavoured by the data. The coherent energy loss model~\cite{Arleo_Peigne_cent_pT_dep_jpsi_suppr_pA_2013} is  not consistent with the data in the two lowest  $p_{\mathrm{T}}$-bins,  but compatible with them at high $p_{\mathrm{T}}$. At backward rapidity, the latter model is in agreement with our data despite some tension at low $p_{\mathrm{T}}$. The calculations by R.~Vogt~\cite{Albaceteetal_predictionspPbatLHC_2013_RVogtversion} for the $p_{\mathrm{T}}$-dependence are consistent with the ALICE results for all three rapidity domains in the provided transverse momentum ranges ($p_{\mathrm{T}}>2.5$ GeV/$c$).

\begin{figure}[t!]
\centering
\begin{minipage}{.5 \textwidth}
\includegraphics[ width=.8 \textwidth]{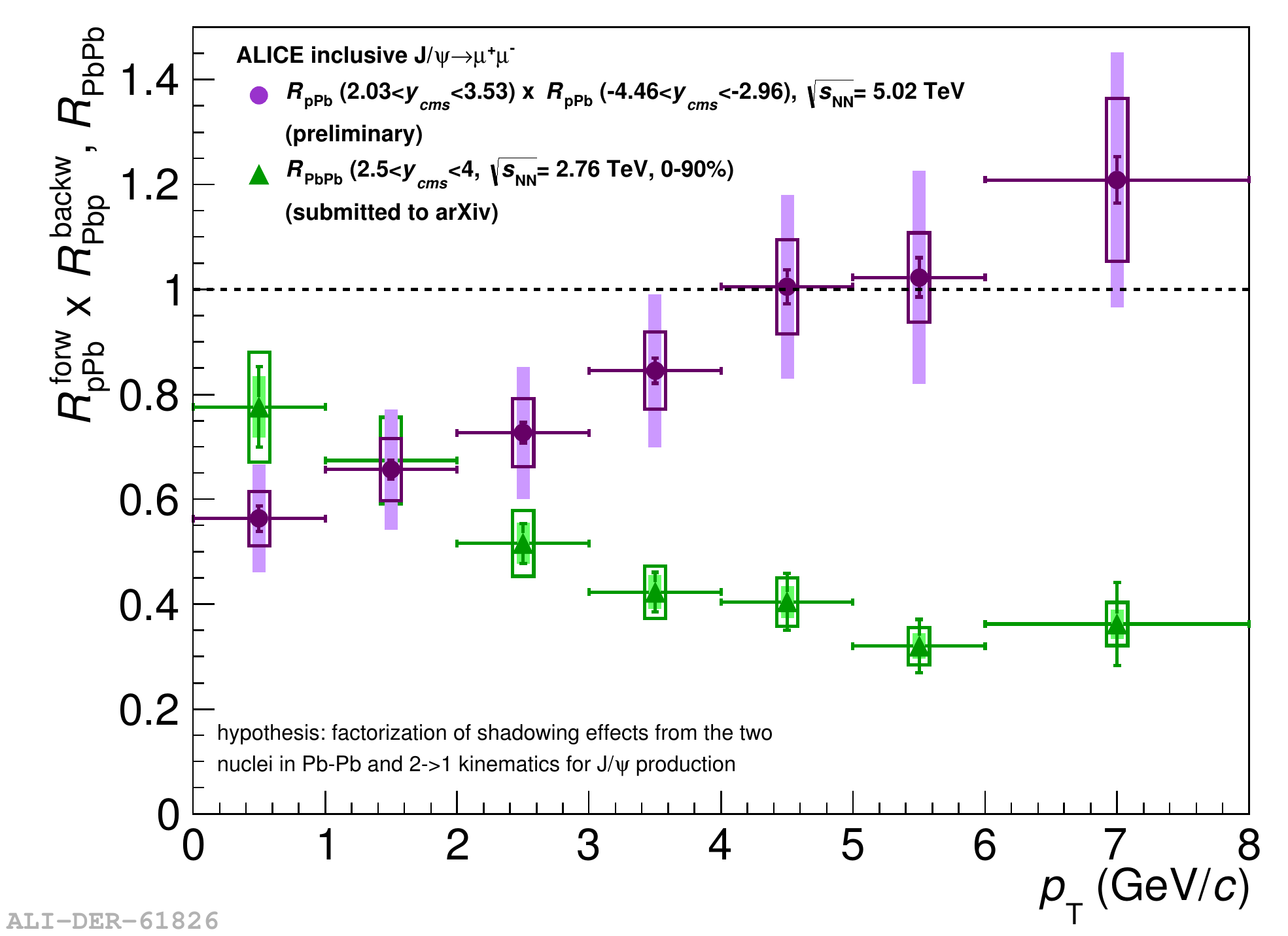}
\end{minipage}%
\begin{minipage}{.5 \textwidth}
\includegraphics[ width=.87 \textwidth]{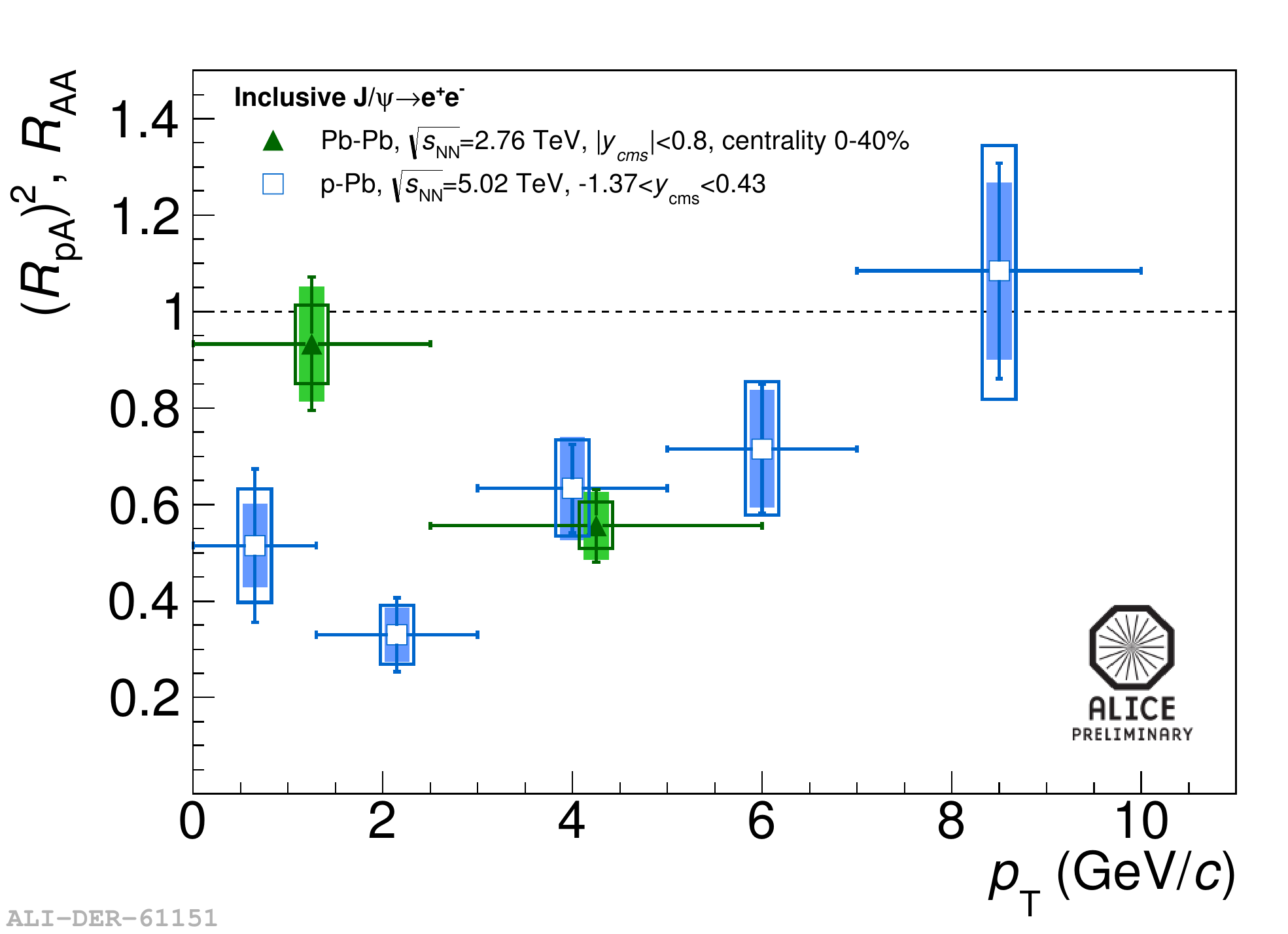}
\end{minipage} 
\caption{The product of $R_{\mathrm{pA,forward}} (p_{\mathrm{T}}) \times R_{\mathrm{pA,backward}} (p_{\mathrm{T}})$ compared to the $R_{\mathrm{AA}}(p_{\mathrm{T}})$ at forward rapidity and the comparison of $R_{\mathrm{pA,midrapidity}}^2$ with $R_{\mathrm{AA,midrapidity}}$. The centrality ranges explored in the AA collision case are not $0-100 \% $ and are indicated in the figures. The display of the uncertainties is the same as for Fig. \ref{fig:rpAasafctofpt}. }
\label{fig:AApAcomp}

\end{figure}

 The Bjorken $x$-values of the lead nucleus probed under a $2 \to 1$ production mechanism assumption ($gg \to J/\psi$) are  approximately matching between the recorded Pb-Pb and p--Pb collision data. Therefore, an expectation for the $R_{\mathrm{AA}}$ based on the $R_{\mathrm{pA}}$ under those kinematic assumptions and the hypothesis of factorization of nuclear effects can be derived by comparing the $R_{\mathrm{pA, backward}} \times R_{\mathrm{pA,forward}}$ ($R_{\mathrm{pA, midrapidity}}^2 $) with $R_{\mathrm{AA,forward}}$ ($R_{\mathrm{AA, midrapidity}}$) assuming that the Bjorken $x$ is  the relevant scaling variable. This comparison is depicted in Fig. \ref{fig:AApAcomp}. It is worth noticing 
 that  there are theoretical models, which do not expect a 'factorization' of cold nuclear matter effects, when one extrapolates from pA to AA collisions at LHC energies~\cite{Kopeliovichetal_notriv_transition_pAtoAA_2011}.
Under the given assumptions, the observed behaviour is in qualitative agreement with expectations from models incorporating non-primordial J/$\psi$ production: in fact, a hint of enhancement of the $R_{\mathrm{AA}}$ result  w.r.t. $R_{\mathrm{pA, backward}} \times R_{\mathrm{pA,forward}}$ ($R_{\mathrm{pA, midrapidity}}^2 $)  at low $p_{\mathrm{T}}$ in the muon spectrometer acceptance(at central rapidity) and a strong suppression at high $p_{\mathrm{T}}$   in the rapidity region explored by the muon spectrometer are visible.

\begin{figure}
\centering
\begin{minipage}{0.49 \textwidth}
\includegraphics[ width=1. \textwidth]{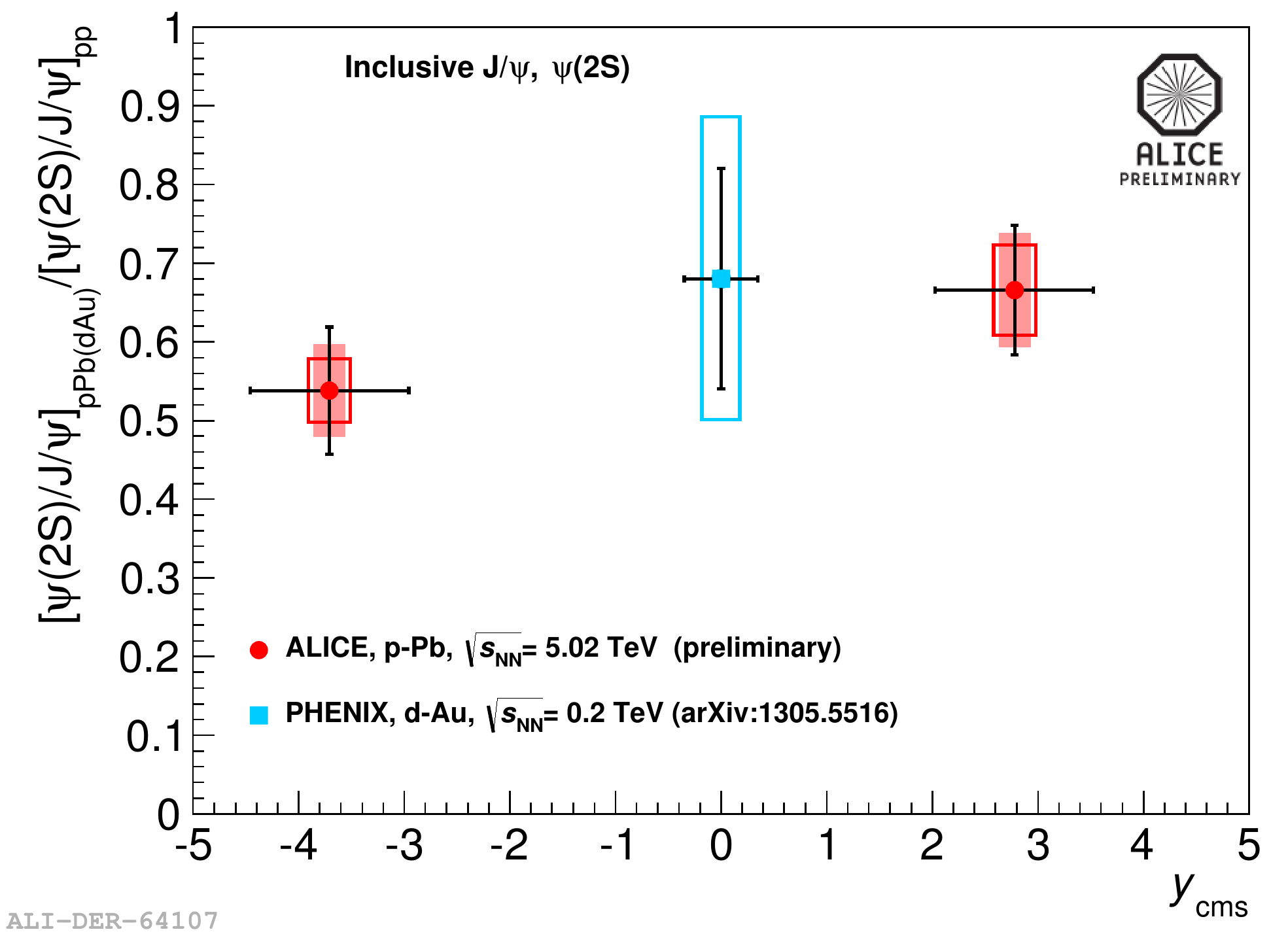}
\end{minipage}
\begin{minipage}{0.45 \textwidth}
\caption{The double ratio $\frac{\sigma^{\psi(2S)}_{\mathrm{pPb}}}{\sigma^{J/\psi}_{\mathrm{pPb}}}/\frac{\sigma^{\psi(2S)}_{\mathrm{pp}}}{\sigma^{J/\psi}_{ \mathrm{pp}}}$ measured by ALICE at $\sqrt{s_{\mathrm{NN}}}= 5.02 $ TeV  in p--Pb collisions and by PHENIX at $\sqrt{s_{\mathrm{NN}}}= 0.2$ TeV in d-Au collisions  is shown in the explored rapidity ranges. Besides the statistical uncertainty depicted as bars, the systematic uncertainty is splitted in case of the ALICE data into a  point-by-point correlated(shaded area) and  point-by-point uncorrelated (boxes) contribution.  
}
\label{fig:psi2s}
\end{minipage}

\end{figure}

A $\psi$(2S)-analysis similar to the J/$\psi$-measurement, is feasible  at forward and backward rapidity only, given the larger statistics.  The result depicted in Fig. \ref{fig:psi2s} unveils a decrease of the inclusive $\psi$(2S) to inclusive J/$\psi$ ratio in p--Pb collisions compared to pp collisions in both studied rapidity domains. A qualitatively similar behaviour was found by PHENIX at central rapidity at $\sqrt{s_{\mathrm{NN} } }$ = 200 GeV in d-Au collisions compared to pp collisions~\cite{Phenix_psiprimechic_jpsi_in_dAu_2013}.
For the efficiency and acceptance corrected yield ratio $\psi$(2S) over J/$\psi$ in pp-collisions, the ALICE result at $\sqrt{s}=$ 7 TeV is used~\cite{ALICE_quarkonia_2014}. 
The energy dependence of the ratio $\sigma^{\psi(2S)}_{\mathrm{pp}}/\sigma^{J/\psi}_{\mathrm{pp}}$ and the slightly different rapidity coverage in pp can introduce a systematic uncertainty. It has been estimated conservatively based on measurements at different collision energies and rapidities. The assigned uncertainty amounts to $3.7\ \%$.
This experimental result is in contradiction with expectations  based solely on shadowing and/or coherent energy loss, since the expected differences between the suppression of J/$\psi$ and $\psi$(2S) are at most of the order $2$-$3 \%$.

In summary, the nuclear modification factor  of J/$\psi$  in p--Pb collisions has been measured by the ALICE collaboration as a function of rapidity and as a function of $p_{\mathrm{T}}$ in the rapidity ranges accessible to the experiment. The results are consistent with shadowing and/or coherent energy loss in case of the rapidity dependence. The coherent energy loss model  and the calculations based solely on shadowing \cite{Albaceteetal_predictionspPbatLHC_2013_RVogtversion} provide a reasonable description of the experimental data at $p_{\mathrm{T}}>$ 2.5 GeV/$c$, whereas the low-$p_{\mathrm{T}}$ nuclear modification factor at forward rapidity are underestimated by the present version of the coherent energy loss model. The inclusive $\psi$(2S)/J$/\psi$ ratio measured in p--Pb collisions shows a clear suppression w.r.t. to the ratio in pp collisions. This behaviour cannot be explained based only on shadowing and/or coherent energy loss.

\bibliographystyle{elsarticle-num}
\bibliography{Literatur}


\end{document}